\documentclass[english,prb,longbibliography,floatfix,superscriptaddress,twocolumn,showpacs,amsmath,amssymb,reprint,latin9]{revtex4-2}
\usepackage[T1]{fontenc}
\usepackage[latin9]{inputenc}
\usepackage{color}
\usepackage{amsmath}
\usepackage{amssymb}
\usepackage{graphicx}

\makeatletter

\newcommand{\lyxmathsym}[1]{\ifmmode\begingroup\def\b@ld{bold}
  \text{\ifx\math@version\b@ld\bfseries\fi#1}\endgroup\else#1\fi}

\PassOptionsToPackage{caption=false}{subfig} 
\usepackage{hyperref}
\hypersetup{
breaklinks=true,
colorlinks=true,
citecolor=blue,
linkcolor=blue,
filecolor=blue,
urlcolor=blue
}
\IfFileExists{lmodern.sty}{\usepackage{lmodern}}{}

\makeatother

\usepackage{babel}
\usepackage{xcolor}

\newcommand{\rd}{\mathrm{d}}
\newcommand{\bS}{\mathbf{S}}
\newcommand{\BB}{\mathrm{BB}}
\newcommand{\ULS}{\mathrm{ULS}}

\newcommand{\su}[1]{\mathfrak{su}(#1)}
\newcommand{\so}[1]{\mathfrak{so}(#1)}
\newcommand{\SU}[1]{\mathrm{SU}(#1)}
\newcommand{\SO}[1]{\mathrm{SO}(#1)}

\begin{document}

\title{The case of SU$(3)$ criticality in spin-2 chains }

\author{Chengshu Li}%
 \email{chengshu@phas.ubc.ca}
 \affiliation{Department of Physics and Astronomy, University of British Columbia, Vancouver BC V6T 1Z1, Canada}
 \affiliation{Stewart Blusson Quantum Matter Institute, University of British Columbia, Vancouver BC V6T 1Z4, Canada}
 
\author{Victor L. Quito}%
 \email{vquito@iastate.edu}
\affiliation{Department of Physics and National High Magnetic Field Laboratory,
Florida State University, Tallahassee, Florida 32306, USA}
\affiliation{Department of Physics and Astronomy, Iowa State University, Ames,
Iowa 50011, USA}

\author{Eduardo Miranda}%
\affiliation{Gleb Wataghin Institute of Physics, The University of Campinas (Unicamp), 13083-859 Campinas, SP,
Brazil}

\author{Rodrigo Pereira}%
 \affiliation{International Institute of Physics and Departamento de F\`{i}sica Te\'{o}rica e Experimental, Universidade Federal do Rio Grande do Norte, 59072-970 Natal-RN, Brazil}

\author{Ian Affleck}%
\affiliation{Department of Physics and Astronomy, University of British Columbia, Vancouver BC V6T 1Z1, Canada}
\affiliation{Stewart Blusson Quantum Matter Institute, University of British Columbia, Vancouver BC V6T 1Z4, Canada}

\author{Pedro L. S. Lopes}
 \email{pedro.lopes@ubc.ca}
\affiliation{Department of Physics and Astronomy, University of British Columbia, Vancouver BC V6T 1Z1, Canada}
\affiliation{Stewart Blusson Quantum Matter Institute, University of British Columbia, Vancouver BC V6T 1Z4, Canada}

\date{\today}

\begin{abstract}
It was proposed in [\href{https://doi.org/10.1103/PhysRevLett.114.145301}{Chen et al., Phys. Rev. Lett. $\mathbf{114}$, 145301 (2015)}] that spin-2 chains display an extended critical phase with enhanced SU$(3)$ symmetry. This hypothesis is highly unexpected for a spin-2 system and, as we argue, would imply an unconventional mechanism for symmetry emergence. Yet, the absence of convenient critical points for renormalization group perturbative expansions, allied with the usual difficulty in the convergence of numerical methods in critical or small-gapped phases, renders the verification of this hypothetical SU$(3)$-symmetric phase a non-trivial matter. By tracing parallels with the well-understood phase diagram of spin-1 chains and searching for signatures robust against finite-size effects, we draw criticism on the existence of this phase. We perform non-Abelian density matrix renormalization group studies of multipolar static correlation function, energy spectrum scaling, single-mode approximation, and entanglement spectrum to shed light on the problem. We determine that the hypothetical SU$(3)$ spin-2 phase is, in fact, dominated by ferro-octupolar correlations and also observe a lack of Luttinger-liquid-like behavior in correlation functions that suggests that is perhaps not critical. We further construct an infinite family of spin-$S$ systems with similar ferro-octupolar-dominated quasi-SU$(3)$-like phenomenology; curiously, we note that the spin-3 version of the problem is located in a subspace of exact G$_2$ symmetry, making this a point of interest for search of Fibonacci topological properties in magnetic systems.
\end{abstract}

\maketitle

\section{Introduction}

\begin{figure}[t]
    \centering
    \includegraphics[width=\columnwidth]{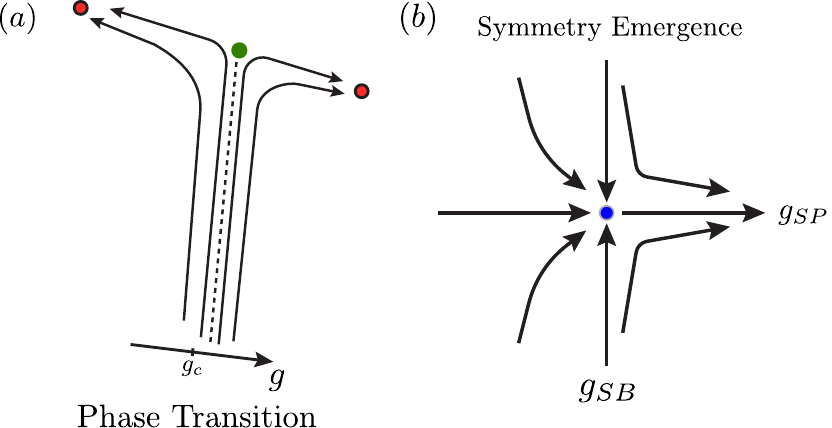}
    \caption{Schematic representation of some typical RG flows: (a) phase-transition at a critical value $g_{c}$ (green point) (b) symmetry emergence. Flows to ``infinite'' parameter values (red points), in general, lead to gaps and stabilization of phases in non-perturbative fixed points.}
    \label{fig:phase_tran}
\end{figure}

The classification of phases of matter remains a cornerstone of condensed matter physics. While the classification by symmetry breaking became a legacy of the field~\cite{landau2013statistical}, more recent discussions have turned attention to cases that go beyond this paradigm, where, in particular, the program of topological classification has had major successes~\cite{Wen_RMP_2017}. A less explored scenario is that of phases where an enhanced symmetry develops in the low-energy sector of a system. This scenario of symmetry emergence and, in particular, of critical (gapless) phases is the focus of this work. As the general phenomenology of these latter situations is not understood, and examples remain few and far between, we will begin by setting the stage. 

When assisted by a renormalization group (RG) picture, standard phase transitions can be understood as follows~\cite{Goldenfeld2018}: one picks a parameter, say $g$, that controls the dynamics of particles in a system at a given length scale. Coarse graining then renormalizes $g$, until it reaches a stable fixed point. This stable fixed point represents a phase where the long-wavelength behavior of the original system has a specific phenomenology. By varying $g$, the flow may find an unstable fixed point, around which any small changes in microscopic choice of the parameter leads to different stable fixed points after RG flow. The system is said to go through a phase transition as $g$ crosses this critical value $g_c$ (which exclusively flows to the unstable critical point). Fig.~\ref{fig:phase_tran} (a) contains a pictorial view of this process. 

However quintessential the story above is, some systems display very different RG flows. An example is displayed on Fig.~\ref{fig:phase_tran}(b). A system may be described by operators that preserve a given microscopic symmetry (with coupling constant $g_{SP}$) and other operators that explicitly break it (with coupling constant $g_{SB}$). If the flow of $g_{SB}$ close to a critical point is always irrelevant, at long wavelengths, the explicit breaking of the microscopic symmetries is not seen. If $g_{SP}$ happens to also be irrelevant in a given direction, a whole critical phase may develop (negative $g_{SP}$ plane in Fig.~\ref{fig:phase_tran}). Besides showing the emergence of scaling symmetry, critical phases participate in a large classes of exotic phenomena including non-Fermi-liquid physics~\cite{Lee_NFL_2018}, topological effects~\footnote{we make reference to Haldane's conjecture in anti-ferromagnetic 1D Heisenberg models~\cite{Haldane1983A,Haldane1983B}}, and Kosterlitz-Thouless type phase transitions~\cite{Goldenfeld2018}.

Here starts the discussion of our explicit physical systems of interest. A well-known example of emergence of highly-symmetric critical phases happens in 1D bilinear-biquadratic spin-1 chains~\cite{Fath1991,Itoi1997}; despite the explicit SU$(2)$ microscopic symmetry of this problem, the existence of an explicit SU$(3)$ symmetric parameter point, aided by an RG analysis as described above, leads to the existence of a critical phase described by an SU$(3)_1$ Wess--Zumino--Witten theory. The existence of this phase is well established, with clear signatures via both analytical calculations and numerical simulations. A refresher on this problem is presented below for self-containing and contrasting purposes.

More recently, a surprising story has been reported in articles~\cite{Chen2012,Chen2015} (henceforth referred to as Chen et al.). There, in the context of bosons in optical lattices, a study was made of general effective models of 1D spin-2 chains close to ferromagnetism. Dimerization and trimerization, and indications of a gapless behavior on the trimerized parameter regime were first reported. Subsequently, a case was made for the existence of an SU$(3)_1$ extended critical phase in the phase diagram of this system. This comes as a surprise: no point of explicit SU$(3)$ symmetry is available in the spin-2 chain parameter space. If an extended SU$(3)_1$ critical phase exists in this model, some novel mechanism must be playing a role, motivating our interest in this problem. We performed extensive analytical attempts at determining some possibility for such mechanism, to no avail. This prompted us to the results of this paper. 

Our purpose in this work is three-fold: (i) to revisit the problem posed by Chen et al., (ii) to situate it in comparison with the well-understood physics of the spin-1 critical SU$(3)_1$ phase, and (iii) to present new numerical results from exact diagonalization as well as from non-Abelian Density Matrix Renormalization Group (DMRG) characterizing the physics of this hypothetical spin-2 SU$(3)_1$ phase, including a critique on whether the previous evidence of Chen et al.~was conclusive. As in previous works, we observe extreme difficulty in the convergence of our routines, rendering the ``gaplessness'' of the phase difficult to determine. We identify static correlation functions of multipolar operators as ideal quantities whose behavior seems to be conclusive independently of finite-size effects. This way, we demonstrate that this hypothetical SU$(3)$ spin-2 critical phase is dominated by ferro-octupolar correlations, in striking contrast with the $2\pi/3$-periodic quadrupolar correlations that dominate the physics of the SU$(3)_1$ Wess--Zumino--Witten theory after logarithmic corrections, as seen explicitly in the spin-1 version of the problem. 

In an attempt to determine an alternate --- to criticality --- origin of our  numerical convergence difficulties, we perform a ``single-mode approximation'' (SMA) study to compare the challenges found in the spin-2 problem with those of a gapless phase proposed to exist close to the ferro-quadrupolar point of the spin-1 bilinear-biquadratic spin chain~\cite{Chubukov1991B,Lauchli2006A}. This comparison proved to still be inconclusive, so we pursued other signatures of critical behavior. Recent years have seen a trend in the use of quantum-information-theoretic tools to study phases of matter. Following this trend, we further the analysis by presenting results on the entanglement spectrum in the phase, and comparing it with universal features expected of gapless systems. 

Besides finite-size scaling, representation-scaling is a paradigm in the analysis of spin chains. This led us to further consider if the phenomenology observed by Chen et al. was an isolated case. Remarkably, we discover a family of models that displays the same tripled-period phenomenology of the spin-2 Hamiltonian but for arbitrary spin-$S$. We demonstrate how these systems are fully quantum and not possible to be described by simple spin-wave theory, and show that for spin-3 our model lies in a region of potential interest to explore G$_2$-symmetry physics in a spin lattice, for which a (G$_2)_1$ critical point/phase could host Fibonacci anyons. 

The paper is organized as follows: in Sec.~\ref{sec:spin-1-recap}, we provide a review and a few complementing results for the spin-1 SU$(3)_1$ phase. In Sec.~\ref{sec:spin-2_recap} we review and extend the known spin-2 model and its phase diagram, including a discussion of the  diverse symmetry-enhanced phases and points, constantly comparing with the spin-1 problem. In Sec.~\ref{sec:spin-2_critique}, we provide most of our new results and analyses on the spin-2 SU$(3)_1$ problem, while in  Sec.~\ref{sec:spin-s} we discuss larger spin-$S$ generalizations. We conclude in Sec.~\ref{sec:conclusion}, with also some suggestions for future research directions. 

\section{Spin-1 SU$(3)_1$ phase review \label{sec:spin-1-recap}}

For contrasting purposes, we begin with a short review of the ``bilinear plus biquadratic'' spin-1 chain phase diagram. The Hamiltonian reads
\begin{align}
    H_{\BB}=\sum_{i}  J\mathbf{S}_i \cdot \mathbf{S}_{i+1}+D \left( \mathbf{S}_i \cdot \mathbf{S}_{i+1} \right)^2 , \label{Eq:BLBQ}
\end{align}
where $\mathbf{S}_i$ are spin-1 matrices on site~$i$. By defining $\tan \gamma \equiv D/J $, the zero-temperature quantum phase diagram of this model can be built by varying the single parameter $\gamma$ in a circle, the radius providing an overall energy scale.

\begin{figure}[t]
    \centering
    \includegraphics[width=\columnwidth]{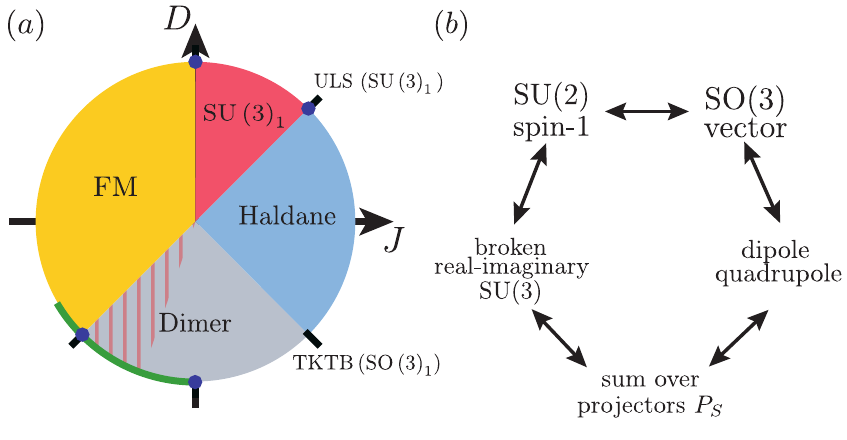}
    \caption{The spin-1 bilinear-biquadratic problem, Eq.~\eqref{Eq:BLBQ}. (a) The zero-temperature phase diagram with the explicit SU$(3)$ points marked in blue. The hashed region corresponds to a hypothetical nematic phase, and the green line is to be matched against the SO(5)-symmetric line in the spin-2 phase diagram below (Fig.~\ref{fig:spin2_pd}). (b) Multiple equivalent ways of representing the SU(2) spin-1 Hamiltonian of Eq.~\eqref{Eq:BLBQ}, listed in Eq.~\eqref{eq:repre}.}
    \label{fig:spin1_pd}
\end{figure}

Four phases exist in the $JD$-circle whose behaviors are well-established in the literature~\cite{Affleck_1986}. Starting at the anti-ferromagnetic Heisenberg model $\gamma=0$ [c.f. Fig~\ref{fig:spin1_pd}(a)] and moving clockwise we have a topological gapped phase (Haldane phase)~\cite{Haldane1983A,Haldane1983B}, a topologically trivial gapped dimer phase~\cite{Barber1989,Klumper1989,Xian1993}, a ferromagnetic phase, and an extended critical phase with period-3 correlations, described by an SU$(3)_1$ Wess--Zumino--Witten conformal field theory (CFT) at low energies~\cite{Fath1991,Itoi1997}. The critical points separating these phases are all very well identified: three of them are points where the lattice problem displays an explicit enhanced SU$(3)$ symmetry ($\gamma=\pi/4,\,\pi/2,\,-3\pi/4$) and one, the so-called Takhtajan--Babujian  point~\cite{Takhtajan1982,Babujian_PhysLetA_1982}~(TKTB, $\gamma=-\pi/4$), is integrable. 

Some debate has been raised regarding the existence of a spin-nematic state as one approaches the SU$(3)$-symmetric ``ferromagnetic'' critical point $\gamma=-3\pi/4$ [hashed red area in Fig~\ref{fig:spin1_pd}(a)]~\cite{Chubukov1991B,Fath1991,Kawashima2002,Ivanov_S1_EFT,Porras2006,Rizzi2005,Bergkvist2006,Lauchli2006A}. A gapped spin-nematic phase was originally proposed for this region by Chubukov~\cite{Chubukov1991B}, but no evidence of gap closing and reopening was found numerically~\cite{Fath1991}. An observed dramatic increase of spin-nematic correlations, as one approaches $\gamma=-3\pi/4$ from the dimerized phase, led to a new critical spin-nematic phase hypothesis for the region~\cite{Lauchli2006A,Porras2006,Rizzi2005}. However, a scenario of a very large crossover, attributed to the influence of the highly-degenerate ferroquadrupolar SU$(3)$-symmetric point on the ground state, has been pointed out as the most plausible explanation to these phenomena~\cite{Lauchli2006A}. We will return to this point when re-evaluating the physics of the spin-2 problem. 

The discussions on multiplets and symmetries above also suggest the relevance of the many different ways of interpreting and rewriting the Hamiltonian in Eq.~\eqref{Eq:BLBQ}. Despite its simplicity, this spin-1 Hamiltonian admits no less than five distinct representations; besides the spin-1 $\SU2$ language in Eq.~\eqref{Eq:BLBQ}, we have:

\begin{equation}
\begin{split}
   H_{\BB}
   = &\sum_{i}J\left(\sum_{a<b}L_{i}^{ab}L_{i+1}^{ab}\right)+D\left(\sum_{a<b}L_{i}^{ab}L_{i+1}^{ab}\right)^{2} \\
    = &\sum_{i}\left[\left(J-\frac{D}{2}\right)\mathbf{S}_{i}\cdot\mathbf{S}_{i+1}+\frac{D}{2}\mathbf{Q}_{i}\cdot\mathbf{Q}_{i+1}+\frac{4D}{3}\right]\\
    = &\sum_{i}\left(-2J+4D\right)\mathcal{P}^i_{0}+\left(-J+D\right)\mathcal{P}^i_{1}
    +\left(J+D\right)\mathcal{P}^i_{2} \\
    = &\sum_{i}\bigg[\left(J-\frac{D}{2}\right)\sum_{a=1}^{3}\Lambda_{i}^{a}\Lambda_{i+1}^{a}+\frac{D}{2}\sum_{a=4}^{8}\Lambda_{i}^{a}\Lambda_{i+1}^{a}\\
    &+\frac{4D}{3}\bigg].\label{eq:repre} 
\end{split}
\end{equation}
In the first line, using the isomorphism of $\su2$ and $\so3$, we write down the Hamiltonian in term of $L_{i}^{ab}$, the three $\SO3$ generators in the vector representation ($a,b=1,2,3$). The second line expresses the Hamiltonian in terms of multipolar operators, with $\mathbf{Q}_i$ the five cartesian quadrupole operators. For the third equality, the Hamiltonian is written in terms of local projectors, with $P^i_{S}\equiv P_{S}\left(\mathbf{S}_{i},\mathbf{S}_{i+1}\right)$ the projector operators of a pair of spins on a total spin $S$ sector. Finally, in the last equality, the $\SU2$ problem is cast as an anisotropic $\SU3$ problem, with $\Lambda_i^a$ the eight generators of the fundamental representation of $\SU3$, broken into two sub-sets of three purely imaginary and five real matrices, using the conventional Gell-Mann basis~\cite{jonesbook}.  The matter of importance is that each of these forms provides an opportunity for insight in the phase diagram and $\SU3$ physics of the spin-1 problem defined by Eq.~\eqref{Eq:BLBQ}.

We thus remark: (i) as observed by~\cite{Tu2008} and discussed below, the phase diagram in Fig.~\ref{fig:spin1_pd}(a), while naively defined for a higher-vector-representation of SU$(2)$, actually finds natural  generalization when seen as a phase-diagram for SO$(N)$-symmetric problems ($N=3$ for $S=1$). This observation alone allows us to extrapolate the phase diagram of Chen et al., as shown below. (ii) By choosing $J=D$, the spin-1 bilinear-biquadratic Hamiltonian is tuned to the integrable Uimin--Lai--Sutherland (ULS) point~\cite{Uimin,Lai1974,Sutherland}. This allows a series of observations regarding the spin-1 SU$(3)_1$ phase, which are relevant in what follows. Applying $J=D$ to the other different representations of the Hamiltonian we obtain:
\begin{equation}
\begin{split}
    H_{\BB}^{\ULS} & = \frac{J}{2}\sum_{i}\left( \mathbf{S}_{i}\cdot\mathbf{S}_{i+1}+\mathbf{Q}_{i}\cdot\mathbf{Q}_{i+1}+\frac{8}{3}\right) \\
    & = 2J \sum_{i}\mathcal{P}^i_{0}+\mathcal{P}^i_{2} = -2J\sum_{i}\left(\mathcal{P}_{1}^{i}-1\right) \label{eq:BBP1} \\
    & = \frac{J}{2}\sum_{i} \left(\mathbf{\Lambda}_i\cdot \mathbf{\Lambda}_{i+1} +\frac{8}{3}\right).
\end{split}
\end{equation}
From the last and first lines, respectively, we remark that the SU$(3)$ invariance is made quite explicit, and that this enlarged symmetry requires identical behavior of the dipolar and quadrupolar fluctuations. From the second line, we observe a suggestive form that will show up again in the spin-2 problem, and will serve as base for further generalizations in Section~\ref{sec:spin-s}.

\begin{figure}[t]
    \centering
    \includegraphics[width=\columnwidth]{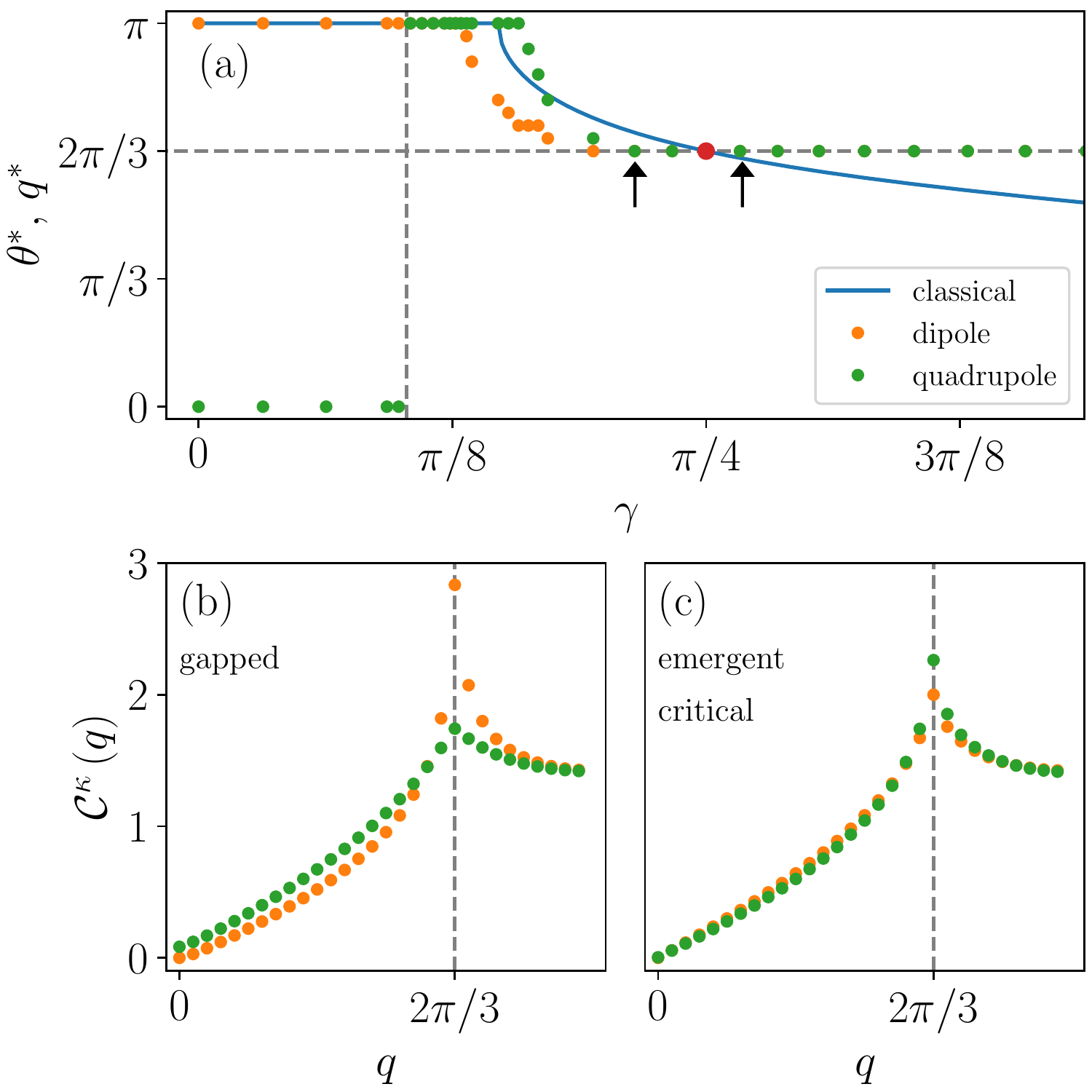}
    \caption{(a) The peak positions $q^*$ of the dipolar and quadrupolar static structure factors, compared to the classical canting angle $\theta^*$. The black arrows correspond to the cuts showed in (b) and (c). The red dot is the the ULS point, $\gamma=\pi/4$, the SU$(3)$ critical point at which the dipolar and quadrupolar correlation functions coincide.  The dominant momentum peak of quadrupolar correlations changes discontinuously at the AKLT point (dashed vertical line). (b,c) The static structure factors around the ULS point in the gapped and emergent SU(3)$_1$ critical, respectively. While, in (c), the system is critical, the emergent nature of the SU(3) symmetry shows by means of finite-size log corrections leading to enhancements of the quadrupolar correlation function as compared to the dipolar one.}
    \label{fig:spin1_theta}
\end{figure}

The establishment of an extended critical phase inheriting properties from the SU$(3)$ point at low energies is the result of a combination of numerical and analytical work~\cite{Nomura1989,Fath1991,Xian1993,Reed1994,Fath1995,Bursill1995,Lauchli2006A}, and is ultimately explained by Itoi and Kato's work via RG~\cite{Itoi1997}. A few more remarks on this region of the spin-1 phase diagram are worthy: it serves as a quintessential example of failure of $1/S$ expansion. As pointed out by Ref.~\cite{Bursill1995}, classically, the spin-wave theory predicts a spiral phase with a continuously varying pitch angle $\theta^*$ as $D/J$ is tuned through the ULS point. Averaging over the azimuthal degeneracy around the canting of successive spins implies that the classical spin-spin correlation functions are given by $\langle \bS_0\cdot\bS_j\rangle=S^2\cos^j\theta^*$\footnote{To see this we can fix $\bS_0$ and use $\langle\bS_j\rangle=\int\prod_{i=1}^j\rd\bS_i P(\bS_i|\bS_{i-1})\bS_j$ and $\int\rd\bS_i P(\bS_i|\bS_{i-1})\bS_i=\cos\theta^*\bS_{i-1}$, where $P(\bS_i|\bS_{i-1})=\frac{1}{2\pi S}\delta\left(S^{-2}\mathbf{S}_{i}\cdot\mathbf{S}_{i-1}-\cos\theta^{*}\right)$ is the conditional probability distribution for $\bS_i$ given $\bS_{i-1}$ fixed.}, peaking at momentum $q=\pi$ for all $\gamma \in [0,\pi/2)$. If spins were restricted to spiral states in a fixed plane, the correlation functions would become $\langle \bS_0\cdot\bS_j\rangle=S^2\cos(j\theta^*)$, with peaks following $\theta^*$ and attaining incommensurate continuous values. This classical large-$S$ description does not correspond to the $S=1$ case of interest: comparing the full line and orange dots in Fig.~\ref{fig:spin1_theta}(a), one sees that the spin-spin quantum correlation function, throughout all $\gamma \in [0,\pi/2)$, is better described by the classical spiral states with spins fixed in a plane. But quantum fluctuations do not only force the spins to remain in a plane, they also saturate what would be the classical canting angle to a commensurate triple periodicity, so that spin-spin correlation functions peak at momentum $q^*=2\pi/3$. Overall, the quantum case is different from both possibilities of classical scenarios; spin-wave and flavor-wave theories only provide good starting points to study this Hamiltonian at specific points of the parameter space, like $\gamma=0$ (anti-ferromagnetic Heisenberg model) or $\gamma=\pi/4$ (SU$(3)$ ULS point) and, still, a full coherent-state quantum treatment is needed to fully capture the physics in these cases.

In Fig.~\ref{fig:spin1_theta}(a), inspired by 
Ref.~\cite{Lauchli2006A}, we further extend the analyses of Ref.~\cite{Bursill1995} and study the static structure factor
\begin{align}
\mathcal{C}^{\kappa}\left(q\right)=\left\langle T_{-m}^{\kappa}\left(-q\right)T_{m}^{\kappa}\left(q\right)\right\rangle, 
\end{align}
where $T_{m}^{\kappa}\left(q\right)$ can be either a dipole ($\kappa=1$, vector) or a quadrupole ($\kappa=2$) operator, the result being independent of $m$ due to SU$(2)$ invariance. This symmetry is incorporated explicitly in our calculations via an SU$(2)$ non-Abelian DMRG routine~\cite{McCulloch2002,Singh2010,Weichselbaum2012,Singh2012}. We evaluate $\mathcal{C}^{\kappa}\left(q\right)$ to find the maximizing momentum $q^*$, and compare it with the classical canting angle $\theta^*$. Besides the results found by Ref.~\cite{Bursill1995}, we observe that the quadrupolar correlations peak at $q^*=0$ before reaching the AKLT point, displaying a ferroquadrupolar uniform behavior. Past the AKLT point, the peak of the quadrupolar fluctuations jumps discontinuously to $q^*=\pi$, similar to the dipolar ones. As one moves towards the ULS point, both static structure factor peaks smoothly interpolate towards $q^*=2\pi/3$, in a crossover behavior associated with the Kosterlitz-Thouless-nature of the transition into the critical SU$(3)_1$ phase. We also note the crossing of the dipolar and quadrupolar $q^*$ curves with the classical $\theta^*$ at the ULS point. While the spin-wave theory fails at this point, this degeneracy lies at the heart of the success of the flavor-wave theory in describing the low-energy theory of this exact SU$(3)$-symmetric point~\cite{LAJKO2017508}. Finally, in Fig.~\ref{fig:spin1_theta}(b) and (c), we display the full static structure factor around $\gamma=\pi/4$; as pointed out by Ref.~\cite{Lauchli2006A}, inside the critical SU$(3)_1$ phase, logarithmic corrections by  SU$(2)$ operators that are irrelevant, in the RG sense, induce an enhancement of the quadrupolar structure factor, in contrast with the dipolar one. Besides that, we call attention to the vanishing tail as $q\to0$, signature of Luttinger-liquid-like behaviour, when $\gamma>\pi/4$. For the Haldane phase, $\gamma<\pi/4$, despite a gap, we remark that $\mathcal{C}^{\kappa=1}\left(q=0\right)=0$ persists for the dipolar correlations, as demanded by  conservation of the total angular momentum and a singlet ground state. It is from the quadrupolar correlations, in fact, that one actually sees the breakdown of Luttinger-liquid behavior via the finite values of $\mathcal{C}$ as $q\to0$~\footnote{In fact, numerically, the $q=0$ quadrupolar correlations remain finite in both the Haldane and critical phases, but it is possible to see that the $y$-intercept increases much more sharply in the Haldane phase; we attribute the finite zero-momentum quadrupolar correlations in the critical phase to the effective nature of the enlarged SU$(3)$ symmetry, and a slow Kosterlitz-Thouless type of renormalization group flow controlling this phase.}.

As a final remark, we note that the conspicuously robust enlarged symmetry of the spin-1 SU$(3)_1$ critical phase does not seem to happen by chance. An algebra isomorphism exists between $\su3_1$ and the integer-spin sector of $\su2_4$~\cite{francesco2012conformal}; since only integer spin multiplets exist in integer spin chains, the SU$(2)$ representation of this problem seems to be carefully appropriate to support an extended SU$(3)$-symmetric critical phase close to the ULS point.

This closes our review and analysis of the spin-1 bilinear-biquadratic problem, and of several of the phenomenological characteristics of its critical SU$(3)_1$ phase. We now turn to describing the spin-2 version of the problem and contrast what is known, and our new results on it, with the spin-1 observations above. Based on the universality of critical theories, we draw conclusions regarding the likelihood of a critical SU$(3)_1$ phase emerging in the spin-2 system.

\section{Spin-2 SU$(3)_1$ phase review} \label{sec:spin-2_recap}

\begin{figure}[t]
    \centering
    \includegraphics[width=\columnwidth]{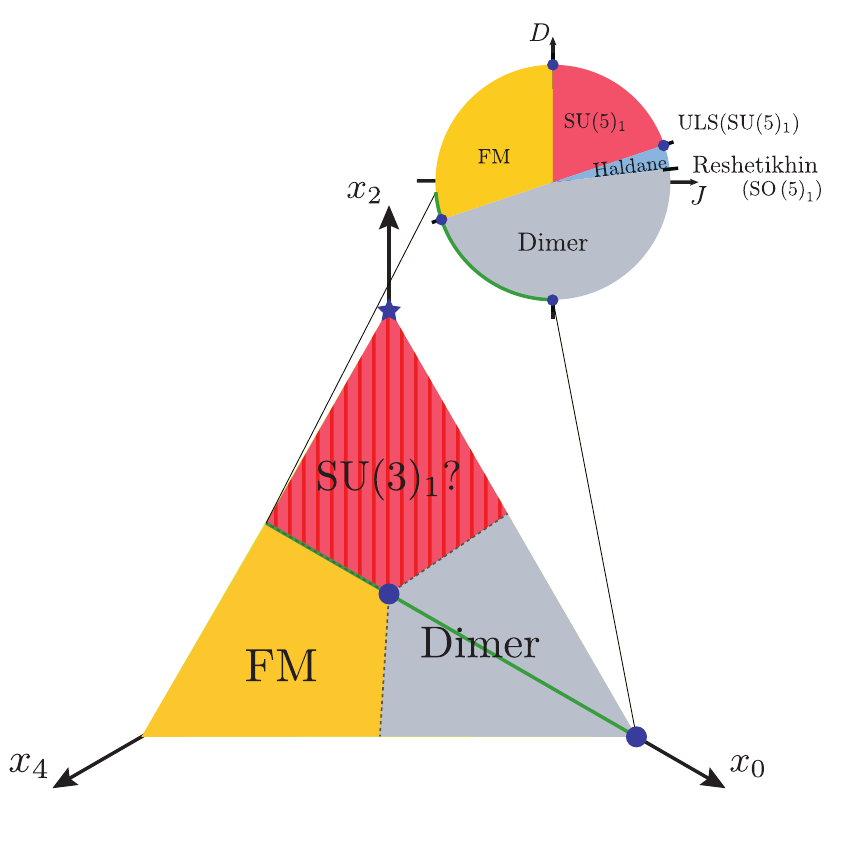}
    \caption{The phase diagram proposed by Chen et al.~\cite{Chen2012} and an extension along an SO$(5)$-symmetric subspace (inset). The grey and yellow phases are dimerized and ferromagnetic, respectively, and the red phases have enhanced symmetry [certainly (uniform) or hypothetically (hashed)]. In the inset, the SU(5) analogue of the spin-1 Uimin--Lai--Sutherland point is located at $\tan^{-1} \frac{1}{3}\approx 18.4^\circ$, and the Takhtajan--Babujian analogue, known as the Reshetikhin point~\cite{Reshetikhin1983,Reshetikhin1985}, is located at $\tan^{-1} \frac{1}{9}\approx 6.3^\circ$).}
    \label{fig:spin2_pd}
\end{figure}

We now move to the most general SU$(2)$-symmetric spin-2 chain, described by the following Hamiltonian~\cite{Chen2012}:
\begin{align}
    H_2 =\sum_{i}\sum_{n=0}^{4}\alpha_{n}\left(\bS_{i}\cdot\bS_{i+1}\right)^{n}
     = \sum_{i}\sum_{n=0}^{4}\epsilon_{n}\mathcal{P}_{n}^{i}.
\end{align}
Here $\alpha_n$ and $\epsilon_n$ are related sets of coupling constants. Motivated by bosonic cold-atomic realizations of this model, Chen et al. considered the phase diagram within the subspace $\epsilon_1=\epsilon_3=0$ and $\epsilon_0,\,\epsilon_2,\,\epsilon_4<0$. Normalizing as $\left(x_{0},x_{2},x_{4}\right)=\left(\epsilon_{0},\epsilon_{2},\epsilon_{4}\right)/\left(\epsilon_{0}+\epsilon_{2}+\epsilon_{4}\right)$, where the denominator represents a global energy scale, the phase diagram is represented by a pyramid built out of a family of triangles with their normal along $(x_0,x_2,x_4)=(1,1,1)$, see Fig.~\ref{fig:spin2_pd}. The distance of a given plane from the origin is the global energy scale.

As shown in Fig.~\ref{fig:spin2_pd}, three phases were identified --- a ferromagnetic phase (FM), a dimerized gapped phase (dimer), and indications of a potentially gapless SU$(3)_1$ phase. Associated with the latter, for finite-sized chains, a tendency towards trimerization was clearly observed via DMRG, together with indications of a gap closing as the system size increases~\cite{Chen2012}. Taking the top of the pyramid in Fig.~\ref{fig:spin2_pd} --- the blue star, henceforth referred to as the pyramidion --- as a representative point deep into the proposed SU$(3)$ critical phase, subsequent exact diagonalization and more DMRG analyses established that the spectrum in this phase has several similarities with the spectrum of the spin-1 chain with $\pi/4<\gamma<\pi/2$. CFT-related quantities were computed including a central charge estimated at $c\approx2$ and conformal towers with scaling dimension $2/3$, both in agreement with the hypothesis of an SU$(3)_1$ critical phase~\cite{Chen2015}. We emphasize that numerical convergence for large-enough system sizes is, however, particularly difficult in this system, implying that these numbers are not necessarily final.

The findings reviewed above come as a surprise: contrasting with the spin-1 problem, no critical point of exact SU$(3)$ symmetry exists here from which a critical phase could extend similarly to the arguments of Itoi and Kato~\cite{Itoi1997}. As already observed by Chen et al., a critical point of larger explicit symmetry here is $x_0=x_2=x_4$, in which the system is SU$(5)$-symmetric, but this is a large-symmetry point with uniform spin orientation, analogous to the $\gamma=-3\pi/4$ point of the spin-1 problem [c.f. Fig.~\ref{fig:spin1_pd}(a)].

In fact, more parallels can be drawn with the spin-1 case. Noting that the $(x_0,x_2,x_4)=(1,0,0)$ point also has exact SU$(5)$ symmetry~\footnote{One needs alternate fundamental and anti-fundamental representations on even/odd sites. See Refs.~\cite{Quito2016,Quito2019} for details.}, we see that the line that bisects the dimer phase in Fig.~\ref{fig:spin2_pd} actually has an enhanced explicit SO$(5)$. To prove so, it suffices to focus on a two-site system and note,
\begin{align}
    &\epsilon_{0}\mathcal{P}_{0}^{i}+\epsilon_{24}(\mathcal{P}_{2}^{i}+\mathcal{P}_{4}^{i}) \nonumber\\
    ={}&J\sum_{a<b}L_{i}^{ab}L_{i+1}^{ab}+D\left(\sum_{a<b}L_{i}^{ab}L_{i+1}^{ab}\right)^{2}+const.
\end{align} where $L_{i}^{ab}$ are the SO(5) generators at site $i$, $J=\epsilon_{24}/2$, and $D=\epsilon_{0}/15+\epsilon_{24}/10$ (as used in Fig.~\ref{fig:spin2_pd}). As observed by Tu et al.~\cite{Tu2008}, the family of SO$(N)$ spin chains with spins in the fundamental (vector) representation have similar phase diagrams. These phase diagrams are one-dimensional and can be arranged in a circle. They follow the pattern of phases and critical points illustrated by the spin-1 (or SO$(3)$) discussion of Section~\ref{sec:spin-1-recap}. This means that, as promised last section, we can extend the regions where we understand the spin-2 phase diagram. We know, for example, that a critical SU$(5)_1$ phase exists in the spin-2 phase space, albeit far from the region in question by Chen et al.  (see the Fig.~\ref{fig:spin2_pd} inset for full details). Back on the ferromagnetic region of the phase space, we note how the green lines in Fig.~\ref{fig:spin2_pd} and its inset match the one depicted in Fig.~\ref{fig:spin1_pd}. In the spin-1 system, we already discussed how the presence of the SU$(3)$ permutation-symmetric critical point $\gamma=-3\pi/4$ induces a large cross-over scale in the dimer phase, leading to a behavior that appears gapless at small system sizes. It is highly suggestive that similar physics could be at play in the spin-2 case, also stretching into the period-three phase to give the impression of that corresponding to a critical phase. Indeed, the role of the competition of ground states induced by adjacent ferromagnetic phase in complicating numerical analysis was observed by Chen et al. 

\section{Spin-2 SU$(3)_1$ phase critique} \label{sec:spin-2_critique}

\subsection{Static structure factor}
\begin{figure}[t]
    \centering
    \includegraphics[width=\columnwidth]{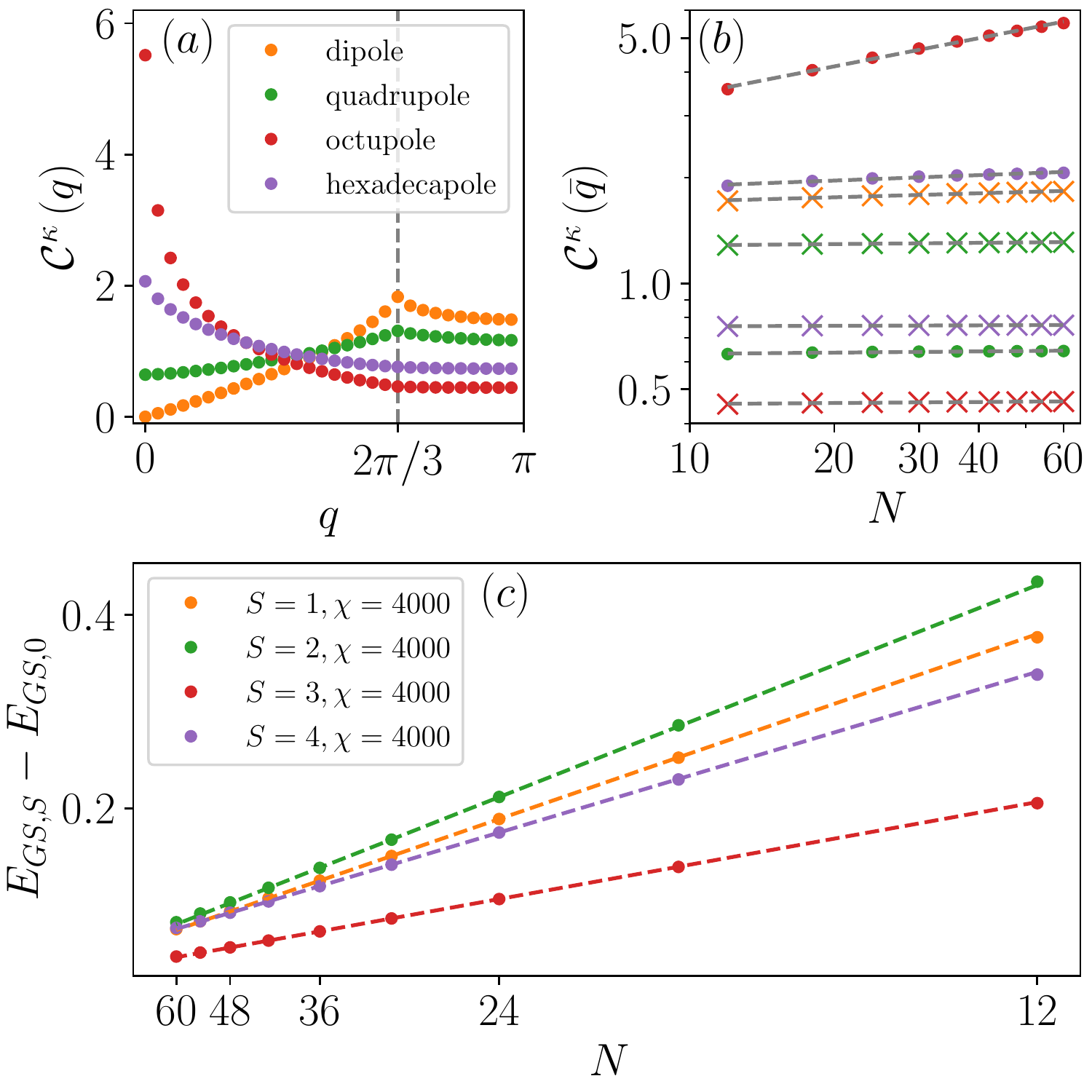}
    \caption{Numerical results on the spin-2 pyramidion Hamiltonian (star point of Fig.~\ref{fig:spin2_pd}). (a) The static structure factors $\mathcal{C}^\kappa(q)$ show a dominance of ferro-octupolar fluctuations, as well as dipole correlations stronger than the quadrupolar ones at $q^*=2\pi/3$, the opposite of what is expected for a critical phase of emergent  SU$(3)$ symmetry (compare with Fig.~\ref{fig:spin1_theta}(b, c)). (b) Finite-size scaling of $\mathcal{C}^\kappa(\bar{q})$ at $\bar{q}=0\,(\mathrm{dots}),\,2\pi/3\, (\mathrm{crosses})$. (c) Finite-size scaling of the energy differences between the lowest-energy spin multiplets $S=1,2,3,4$ and the singlet ground state. We chose $N$ to be multiples of six and spaced the values according to $1/N$ as $E_{GS,S}-E_{GS,0}$ is expected to scale as $1/N$.}
    \label{fig:spin2_corr}
\end{figure}

To shed light on the hypothetical SU$(3)_1$ spin-2 phase, we draw inspiration from the spin-1 results above. We consider the multipolar static correlation functions $\mathcal{C}^\kappa(q)$ at the spin-2 pyramidion point. Due to the higher-spin representation, not only dipolar ($\kappa=1$) and quadrupolar ($\kappa=2$) correlations are available, but octupolar and hexadecapolar are too ($\kappa=3,\,4$, respectively). Again, we obtain our results by performing SU(2) non-Abelian DMRG calculations and the results are shown in Fig.~\ref{fig:spin2_corr}(a). Remarkable differences are found between the well-established SU$(3)_1$ spin-1 phase and the spin-2 case. While for spin-2 dipole and quadrupole fluctuations are indeed dominated by triple-periodicity, i.e. $q^*=2\pi/3$, the quadrupole fluctuations do not vanish as $q\to0$, as would have been expected for critical behavior. Furthermore, the dipole correlations are stronger than the quadrupolar ones, the opposite of what is argued to be expected from an  SU$(3)$ critical phase emerging of an SU$(2)$ lattice problem~\cite{Lauchli2006A,Itoi1997}. At last, we verify that octupolar and hexadecapolar correlations are dominated by a uniform behavior, i.e. $q^*=0$, and these correlations are also stronger than the dipolar and quadrupolar ones. A ferro-octupolar phase is more aligned with the physics than a critical SU$(3)_1$ one. 

Chen et al. observed via exact diagonalization that, indeed, the ordering of the lowest energy level for each spin multiplet in the spin-2 pyramidion point is not the same as that of the emergent SU(3) spin-1 phase (in ascending order, $S=0,\,3,\,1,\,4,\,2$, for the former, $S=0,\,2,\,1,\,3,\,4$, for the latter). They claimed that the order of these multiplets was a result of finite-size effects together with the fact that the SU(3) invariance is emergent, not exact. To verify this hypothesis, we performed a finite-size scaling analysis of the correlation functions, following its values at $\bar{q}=0,\,2\pi/3$, as well as of the energy of the ground state of each spin $S=1, 2, 3, 4$ relative to the singlet ground state. We choose system sizes in multiples of 6 to avoid two- and three-fold frustration under periodic boundary conditions, with values  between $N=12$ and $N=60$. The results are shown in Fig.~\ref{fig:spin2_corr}(b). Despite not going to system sizes as large as Chen et al., we push the limits on bond dimensions beyond what they considered, favoring precision over system size (more details are given in Appendix~\ref{app:num}). Our findings indicate that the ferro-octupolar correlations tend to increase with system size most intensely, followed by the ferro-hexadecapolar ones. The period-3 dipolar and quadrupolar correlations seem to show little to no appreciable increase as functions of the system size. As for the energy levels, we find, as shown in Fig.~\ref{fig:spin2_corr}(c), that indeed a tendency towards an exchange of order of the lowest-energy multiplets cannot be discarded. We plot the energy differences for different $N$ by scaling the x-axis as $1/N$ as, in this way, the relation is expected to be linear~\cite{Itoi1997}. Yet, we emphasize that, despite our efforts to push the limits of bond dimension in our system --- $\chi=4000$, corresponding to $\chi_{ns}\sim 35000$ of non-symmetric matrix-product states --- we cannot claim full convergence for system sizes $N\gtrsim 48$~\footnote{Also, contrast with Chen et al. who pushed system sizes to 120, but with bond dimensions truncated at max $\chi=2800$.} (again, more details are shown in Appendix~\ref{app:num}). 

\subsection{Single-mode approximation}

\begin{figure}[t]
    \centering
    \includegraphics[width=\columnwidth]{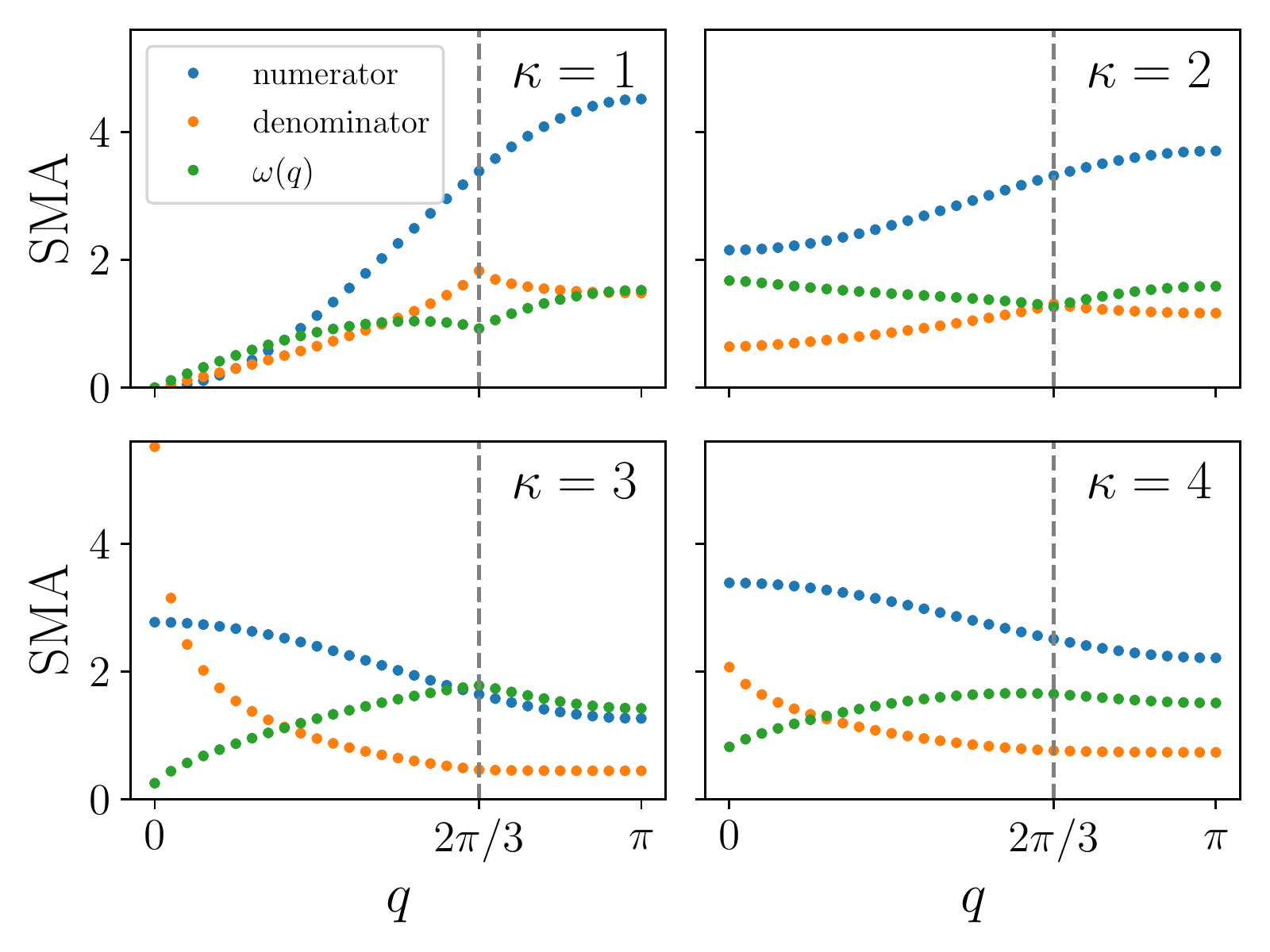}
    \caption{The single-mode approximation. We plot $\omega_{q}$ as well as the numerator $\langle 0|[(O(q))^{\dagger},[H,O(q)]]|0\rangle$ and the denominator $\langle O(q)|O(q)\rangle$ of $\omega_{q}$ separately [see Eq.~\eqref{eq:sma}] with the choice $O(q)=T^\kappa_m(q)$. All calculations have system size $N=60$ and bond dimension $\chi=4000$.}
    \label{fig:SMA}
\end{figure}

Being unable to fully verify the signatures of a critical phase, we
return to the problem of the apparent criticality in the region between $\gamma=-\pi/4$ and $-3\pi/4$ of the spin-1 problem. Inspired by that, we contemplate the possibility of the a large crossover scale due to the closeness of the high-symmetry ferromagnetic region. To address that, we follow Ref.~\cite{Lauchli2006A}, which suggested the use of the ``single-mode approximation'' (SMA) as a venue to verify such a scenario. 
The SMA process consists in constructing momentum-specific trial states
\begin{equation}
    \left|O\left(q\right)\right\rangle =O\left(q\right)\left|0\right\rangle, 
\end{equation}
for a given operator $O\left(q\right)$ carrying quantum numbers of a sector of interest in the Hilbert space, and $\left|0\right\rangle $ is the ground state, which we find via DMRG. The mean value of the Hamiltonian in this state may be written as
\begin{align}
    \omega_{q}&=\frac{\left\langle O\left(q\right)\left|H\right|O\left(q\right)\right\rangle }{\left\langle O\left(q\right)\left|O\left(q\right)\right.\right\rangle} -\langle0|H|0\rangle\\
    &=\frac{1}{2}\frac{\left\langle 0 \left|\left[\left(O\left(q\right)\right)^{\dagger},\left[H,O\left(q\right)\right]\right]\right|0\right\rangle }{\left\langle O\left(q\right)\left|O\left(q\right)\right.\right\rangle } \label{eq:sma}
\end{align}
and serves, like any variational energy, as a strict upper bound on the gap for the sector of interest~\footnote{For this to work it is necessary that $\langle0|O(q)\rangle=0$, which can be easily checked. See e.g. Sec. 6.2 of \cite{Tasaki2020}.}. The power of this method comes from the fact that this upper bound can be seen to vanish both from the perspective of (i) the divergence of the denominator, typical for critical phases or phase transitions, as well as from (ii) the vanishing of the numerator, which requires a commutation of $O(q)$ and the Hamiltonian. In the latter case, an anomalously small upper bound to a system gap can emerge, giving the impression of critical behavior. In the spin-1 chain study, the latter case was realized by noting that at $q=0$ the quadrupole operator $T^2_m(q)$ commuted with the Hamiltonian at $\gamma=-3\pi/4$~\cite{Lauchli2006A}. 

We thus ask ourselves if, along the lines of scenario (ii), the presence of the SU$(5)$ point and the SO$(5)$ ``ferromagnetic'' line in the phase diagram of the spin-2 chain (c.f. Fig.~\ref{fig:spin2_pd}) could be influencing the apparent critical-like physics at the pyramidion. We compute $\omega_q$ for the operators $O\left(q\right)=T_m^\kappa(q)$ to estimate the upper energy bounds for each $\kappa$ sector, corresponding to the total angular momentum $S$ sectors. The results are presented in Fig.~\ref{fig:SMA}. We find that the vanishing of the numerator in Eq.~\eqref{eq:sma} only happens for the dipolar $\kappa=1$ sector, at $q=0$. This result is trivial, as the total angular momentum is naturally preserved due to the exact SU$(2)$ symmetry of the problem; the SMA loses predictive capacity in this angular and linear momentum sector. Another feature we remark happens in the case of $\kappa=3$, the octupolar sector, which displays $\omega_q\to0$ as $q\to0$ due to the divergence of the structure factor in the denominator of Eq.~\eqref{eq:sma}. This, together with the other notable points of minima for $\omega_q$ (namely, $q=0$ for $\kappa=4$ and $q=2\pi/3$ for $\kappa=1$ and $2$), is more closely related to scenario (i) described above and cannot offer predictions beyond those of the finite-size scaling performed in the previous sub-section. 

In conclusion, the SMA analysis does not support nor contradict claims of critical behavior in the spin-2 problem. If we moved away from the pyramidion towards the ferromagnetic phase, the SO$(5)$-symmetric phase boundary certainly leads to vanishing SMA upper bounds in all sectors we considered at $q=0$, but the pyramidion point itself is far enough for the commutators in the numerator of $\omega_q$ to be comfortably finite. Reversing the point of view, the opposite conclusion may bring some extra value: in Ref.~\cite{Chen2015}, it was claimed that energy-level considerations were more trustworthy than entanglement-entropy scaling analyses when trying to extract CFT characteristics, albeit the latter being more typically exploited for these purposes. Indeed, while our SMA results do not capture entanglement effects, they do indicate no anomalous effects in the energy levels for the Hamiltonian at the pyramidion point, so that energy-level scaling analyses seem to be limited just by regular finite-size scaling convergence. 

\subsection{Entanglement spectrum}

A further tool that has recently been used to classify phases of matter, as well as analyse criticality, is entanglement. In 1D criticality, a commonly studied quantity is the entanglement entropy, known to be directly related to a conformal field theory invariant, namely the central charge~\cite{EE_crit}. More recently, it has been shown by Calabrese and Lefevre that not only the central charge is a specific signature of critical points and phases, but the whole distribution of eigenmodes of the reduced density matrix takes a special shape~\cite{Calabrese2008}. The predictions of Calabrese and Lefevre were verified in numerical simulations~\cite{Pollmann2010}. Chen et al. already extensively verified the entanglement entropy scaling, finding, as mentioned, that robust CFT information was hard to obtain. Here we extend the entanglement study of the pyramidion spin-2 Hamiltonian and contemplate the entanglement spectrum predictions of Calabrese and Lefevre in another attempt to solve the deadlock between presence or absence of criticality.

One considers the reduced density matrix of a subsystem $A$, decomposed in terms of eigenmodes as
\begin{align}
    \rho_A=\sum_i \lambda_i\left|\lambda_{i}\right\rangle \left\langle \lambda_{i}\right|, \label{eq:Efunct}
\end{align}
$\lambda_i$ being the eigenvalues. From these, a probability distribution can be built for the eigenvalues, $P(\lambda)$, for which the mean number of eigenvalues larger than a value $\omega$ assume a very simple form in conformal invariant systems, namely
\begin{align}
    n(\omega)=\int_{\omega}^{\omega_\mathrm{max}} d\lambda P(\lambda) =I_0\left(2\sqrt{b \log{\frac{\omega_\mathrm{max}}{\omega}}}\right), \label{eq:Efit}
\end{align}
where $I_0$ is the zeroth order modified Bessel function of the first kind, $\omega_\mathrm{max}$ is the largest eigenvalue of $\rho_A$ and is also related to the parameter $b$ according to
\begin{align}
    b=-\log{\omega_\mathrm{max}}=\frac{c}{6}\log{L_\mathrm{eff}} \label{eq:btoc},
\end{align}
where $c$ is the conformal central charge, $L_\mathrm{eff}=N\sin{\pi \ell_A/N}$, with $N$ the system size, and $\ell_A$ the size of the sub-region $A$~\cite{Calabrese2008}. We always take $\ell=N/2$, so that $L_\mathrm{eff}=N$.  

Thus, constructing $n(\omega)$ is, at first sight, tantamount to measuring the central charge $c$, a quantity that has already been considered by Chen et al. via the entanglement entropy with several convergence limitations. Yet, $n(\omega)$ goes beyond that and displays a very specific functional form for conformal systems [Eq.~\eqref{eq:Efit}] that deserves being considered on its own as a potential signature of criticality.

\begin{figure}
    \centering
    \includegraphics[width=\columnwidth]{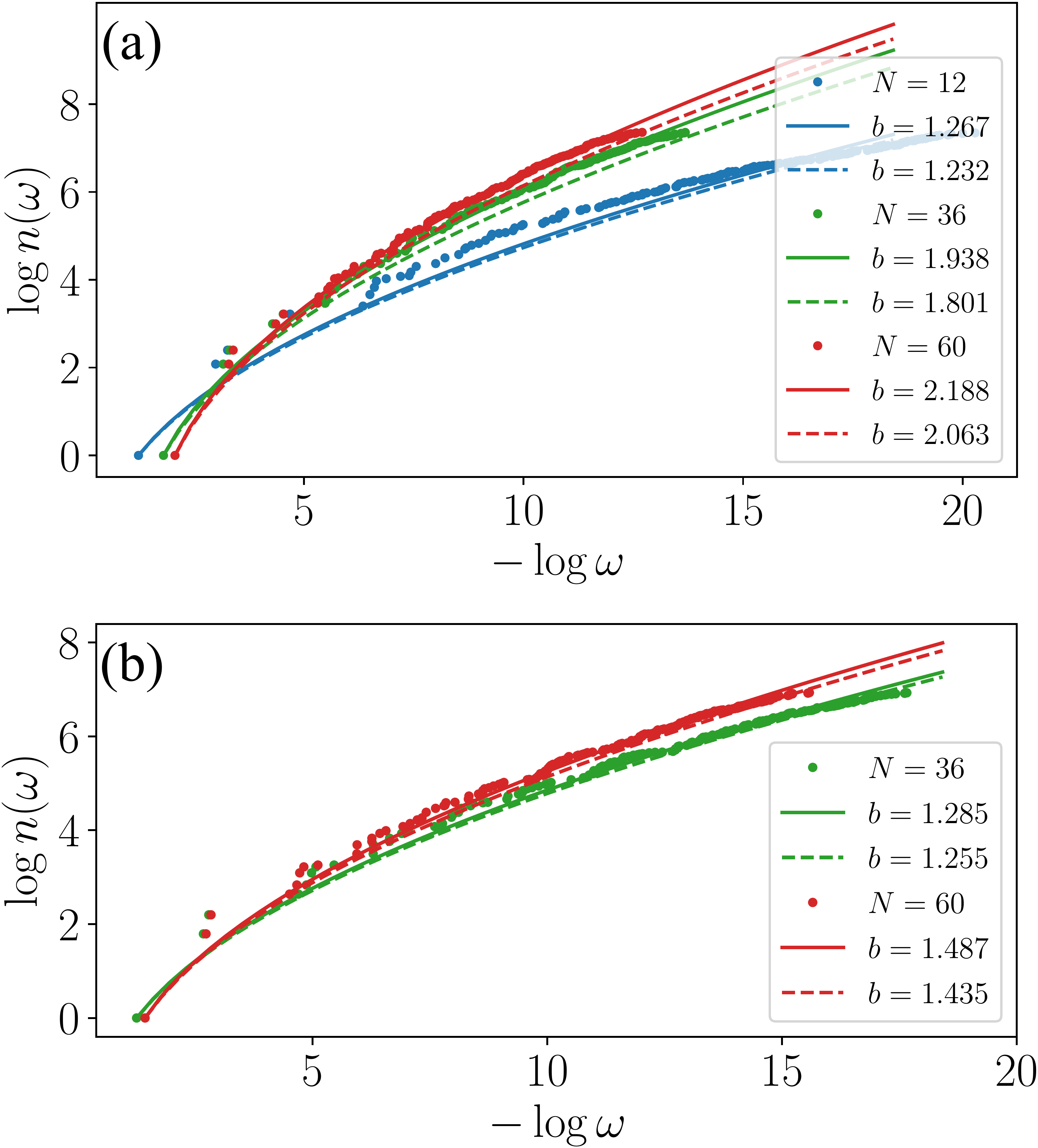}
    \caption{The cumulative distribution function of the entanglement spectrum $n(\omega)$ of a subsystem of half the system size for (a) the spin-2 model and (b) the spin-1 model with $\tan\gamma=1.1$. We also show the Calabrese-Lefevre functional form with fitted $b$ (full) and the analytical $b=-\log\omega_\mathrm{max}$ (dashed). }
    \label{fig:Espect}
\end{figure}

In Fig.~\ref{fig:Espect}(a) we display the numerical data for $n(\omega)$ for a system sizes $N=12,\,36$ and $60$, and consider it against the right hand side of Eq.~\eqref{eq:Efit}. We consider both fitting the data for the functional form of $I_0$ (full lines), as well as a fixed curves with the analytical form of $b$ (dashed curves). As we see, as the system size increases, the functional form of $I_0$ fits numerical data better and better. Yet, discrepancies arise between the fitted $b$ and the analytical prediction of $b$. The discrepancies are smaller for small system and become more apparent as the system size increases from 12 to 24 (as we verified). As the system size continues increasing to 60, the discrepancy tends to diminish again, albeit more slowly. The fitted values of $b$ are consistently smaller than their analytical counterparts and, from Eq.~\eqref{eq:btoc}, result in estimates for the central charge $c$ ranging around $2.9$ to $3.2$. These estimates seem to be consistent with Chen et al. for similar system sizes (although, again, we pursue larger truncation dimensions). For contrasting purposes, we also perform the same calculation in the spin-1 SU(3)$_1$ critical phase (Fig.~\ref{fig:Espect}(b)). We choose the point $\tan\gamma=1.1$ slightly off the ULS point, where the symmetry is emergent instead of exact. In this case, the fitted $b$ its analytical predicted value agree better, and both give an estimate of the central charge $c\simeq2.1-2.2$ closer to the expected $c_{SU(3)_1}=2$.

\section{spin-$S$ and beyond} \label{sec:spin-s}

Our results cast doubt in the existence of the spin-2 SU$(3)_1$ critical phase. Nevertheless, an interesting realization prompts us to further consider this possibility. Rewriting the spin-2 pyramidion Hamiltonian in a projector language, it reads
\begin{align}
    H_{2}^{\star}=-\sum_{i}\mathcal{P}_{2}^{i}.
\end{align}
We remark on the similarity between this and Eq.~\eqref{eq:BBP1}. As it turns out, for an arbitrarily chosen \emph{integer} spin $S$, the ground state of a three-site problem with Hamiltonian $\sim -(\mathcal{P}_S^1+\mathcal{P}_S^2)$ is always a singlet. For larger system sizes, quantum fluctuations scramble and entangle these 3-spin singlets, generating an SU$(3)_1$ critical phase for spin-1 and, as was hypothetically proposed, also for spin-2. This observation leads to an irresistible idea: could potentially a family of SU$(3)_1$ critical phases exist for \emph{arbitrary integer} spin-$S$ systems? (Contrasting with the family of SU$(2)_1$ phases that exist for \emph{half-integer} spin chains, driven by Haldane's topological arguments in the anti-ferromagnetic Heisenberg models.) We leverage our learning so far to explore this question.

\begin{figure}
    \centering
    \includegraphics[width=\columnwidth]{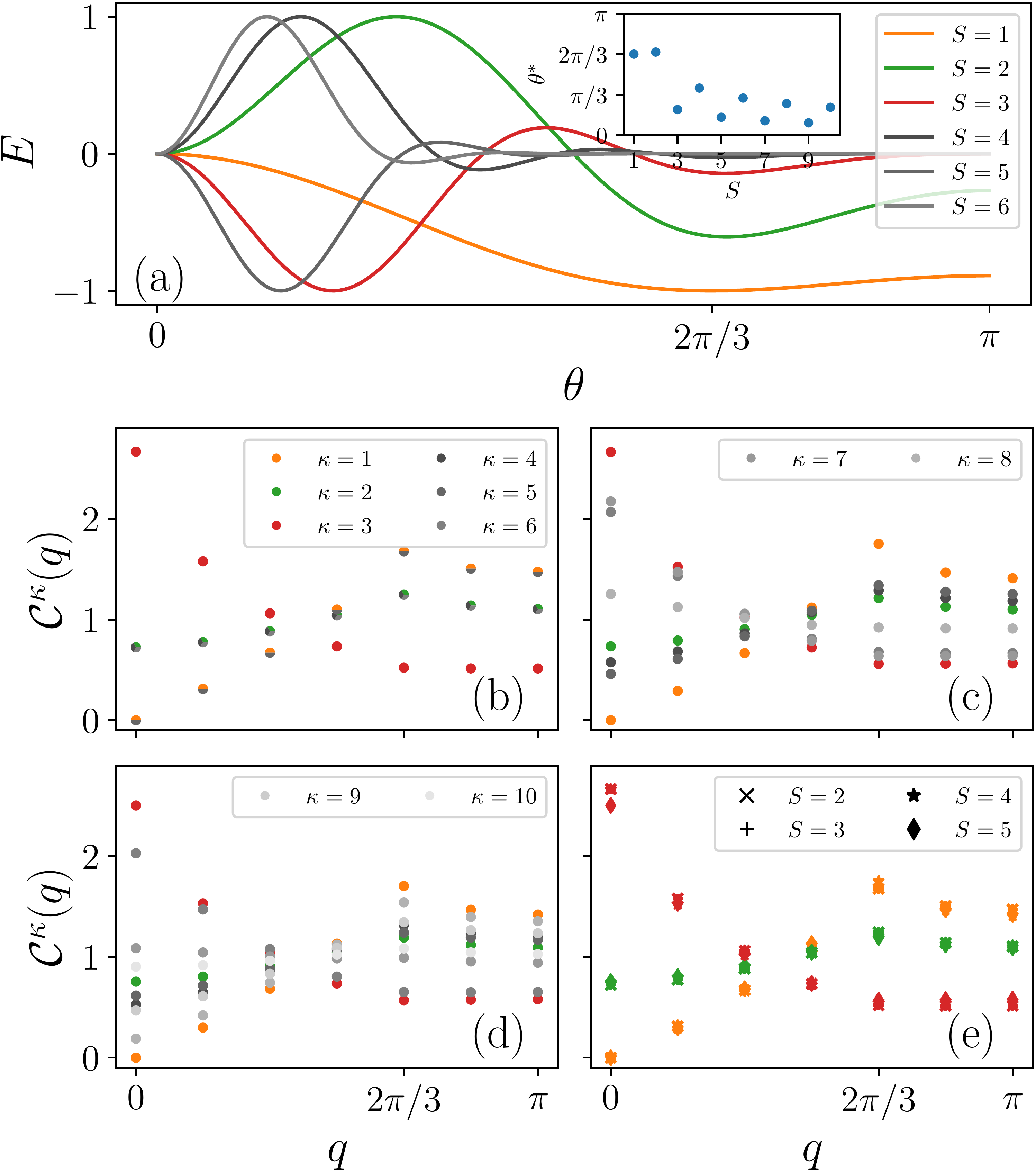}
    \caption{(a) The classical energy $E(\theta)$ associated with the spin-$S$ Hamiltonian $H=-P_S$, with $P$ the projector onto the multiplet of total angular momentum $S$, a possible extrapolation of the spin-2 Hamiltonian. Inset: the angle $\theta^*$ that mininizes the energy. We have rescaled $E(\theta)$ to the interval $-1$ to $1$ to better present the trend. For $S=1,2$, the peak is at $2\pi/3$ but for higher spins, the position shifts towards smaller values. (b-d) The static structure factor for the quantum Hamiltonian $H=-P_S$ for $S=3,4,5$ for different values of $\kappa$ and system size $N=12$. We use the same color code as in (a), but the meaning is distinct: each panel (b-d) corresponds to a different spin value, and the color code refers to the values of $\kappa$. For $S\geq4$ (a) or $\kappa\geq4$ (b-d), the points are shown in ever lighter shades of grey. For $S=3$, an extra $G_2$ symmetry is reveled:  $\kappa=1$ (dipolar) and $\kappa=5$ (triacontadipolar) have the correlation, as well as $\kappa=2,4,6$. (e) A comparison of all the commonly available $\mathcal{C}^\kappa$ for chains with $S=2,3,4,5$. We note that their behavior of correlations is spin-independent, independently also of finite-size corrections.
    }
    \label{fig:Sscaling}
\end{figure}

We consider an spin-$S$ SU$(2)$-invariant chain with Hamiltonian
\begin{align}
    H_{S}=-\sum_{i}\mathcal{P}_{S}^{i}. \label{eq:spinS}
\end{align}
We start by studying the classical spin texture expected for this Hamiltonian. We write the projection operator explicitly so that
\begin{equation}
H_S=-\sum_i\prod_{k\neq S}\frac{\mathbf{J}_{i,i+1}^2-k(k+1)}{S(S+1)-k(k+1)},
\end{equation}
where 
\begin{equation}
\mathbf{J}_{i,i+1}^2=\mathbf{S}_i^2+\mathbf{S}_{i+1}^2+2 \mathbf{S}_i\cdot \mathbf{S}_{i+1},
\end{equation}
and substitute $\mathbf{J}_{i,i+1}^2=2S(S+1)+2S^2\cos\theta$, with a uniform first-neighbor canting angle $\theta$. This results in a classical energy given by
\begin{equation}
E(\theta)=-N\prod_{k\neq S}\frac{2S(S+1)+2S^2\cos\theta-k(k+1)}{S(S+1)-k(k+1)},
\end{equation}
and we search the favored ground state by finding the angle $\theta^*$ that minimizes the energy. The results  are plotted in Fig.~\ref{fig:Sscaling}(a). As we know, for $S=1$ we have $\theta^*=2\pi/3$ exactly. While for $S=2$, $\theta^*$ is shifted from $2\pi/3$ only mildly, for larger spin $S$ we find strong deviations from this triply-periodic value. 

Naturally, this serves as another demonstration of the failure of the classical description of spin waves in 1D systems. The scrambling of 3-site singlets indicate that correlations of arbitrary $q^*$ are not possible and will not match the classical $\theta^*$. To verify this is the case, we again use our non-Abelian DMRG code to compute the static structure factor for all available higher-order tensor operators extending the spin-1 and spin-2 results to $S=3,4,5$. While the DMRG convergence gets penalized as $S$ increases, our experience is that the position of the dominant momenta in $\mathcal{C}^\kappa(q)$ is largely stable and well-fixed even at system sizes as small as $N=12$.

Figs.~\ref{fig:Sscaling}(b-d) display the results for $S=3,\,4$ and $5$ and are representative of the trend we see for all larger $S$. Correlations peak sharply at either $q^*=0$ or $2\pi/3$. As we find, uniform octupolar correlations are \emph{always} the dominating ones, and dipole and quadrupole correlations always peak together at $2\pi/3$ (with dipoles dominating over quadrupoles, just as in $S=2$). Higher multi-polar correlation functions are also available for higher spin chains, but are not the center of our attention; they are displayed in grey in the figure. In Fig.~\ref{fig:Sscaling}(e), we compare $\mathcal{C}^\kappa(q)$ with $\kappa=1,\,2,\,3,\,4$ for spins $S=2$ to $5$, and note that, independently of finite-size effects, the behavior of the correlation functions are the same for all representations.

A ``surprise'' can also be observed: for $S=3$ (Fig.~\ref{fig:Sscaling}(b)), we observe an exact degeneracy between dipolar ($\kappa=1$) and triacontadipolar (32-polar) ($\kappa=5$) operators (orange and shades of grey), as well as between quadrupolar ($\kappa=2$) , hexadecapolar ($\kappa=4$) and hexacontatetrapolar (64-polar) ($\kappa=6$) operators (green and shades of grey). These two sets, together with the distinct octopolar correlations ($\kappa=3$) (red), lead to multiplets $7+14+27$. Combining with the  $\kappa=0$ singlet then reproduces the levels of the group G$_2$, hinting to an exact microscopic G$_2$ symmetry of this model for $S=3$.  We list in Appendix~\ref{app:g2} the generators of its corresponding Lie algebra, $\mathfrak{g}_2$ and an analysis of the Hamiltonian. Upon a careful look, the degeneracies should not be a surprise at all: we demonstrate rigorously that a region of exact G$_2$ symmetry exists inside the phase space of the $S=3$ rotational-invariant Hamiltonians (also discussed in Appendix~\ref{app:g2}). We thus conclude this section with a comment on future directions of inquiry: in a similar spirit to our previous searches of symmetry emergence, this observation makes the $S=3$ spin chain a relevant system to search for $(\mathrm{G}_2)_1$ critical points and phases, of importance in the search of Fibonacci anyons, in fine-tuned region of the parameter space of an SU(2)-invariant system. Whether one can find robust phases where the larger symmetry is manifest remains a problem for the future.

\section{Conclusion} \label{sec:conclusion}

We re-evaluated the signatures of SU$(3)_1$ criticality in integer-spin 1D chains, in particular in the spin-2 system of Refs.~\cite{Chen2012,Chen2015}. Comparing with the well-known case of spin-1, we analyzed, via SU$(2)$-symmetric non-Abelian DMRG, the static structure factors of multipole tensor operators in the spin-2 problem as profoundly in the hypothetical SU$(3)_1$ phase as we could. Our results indicate domination of ferro-octupolar correlations, over $2\pi/3$-periodic dipolar and quadrupolar, as well as missing signatures of long-range Luttinger-liquid-like behavior in the quadrupolar correlations. We push the convergence of finite-size scale analysis in comparison with previous literature results and, despite the difficulty in attaining full convergence, trends indicate that the results above should not change in the thermodynamic limit but rather become more pronounced. 

By tracing parallels and performing symmetry analyses, we also extended our understanding of phases in the spin-2 parameter space, pointing to a phase of extended SU$(5)_1$ symmetry, and showing that the problem proposed by Chen et al. lies in a similar place with an anomalously critical-looking region of the spin-1 phase diagram. As done previously in the case of spin-1~\cite{Lauchli2006A}, we used a single-mode approximation to explore the proximity of high-symmetry uniformly ordered critical lines and a ferromagnetic phase as explanations for the apparent spin-2 critical behavior. While we found little evidence that this is the actual origin of the problem, our results indicate that the energetics and correlation analysis performed are trustworthy and robust. 

We also explored the entanglement spectrum of the spin-2 problem as a potential venue to determine whether this Hamiltonian is indeed critical or not. The distribution of eigenvalues is well fitted by analytical laws valid in conformal systems. Yet, we verify discrepancies in the fitting parameter and difficulty in estimating the central charge as previously reported in the literature. That the critical form for the distribution of eigenvalues of the entanglement spectrum is well respected is, perhaps, not surprising, as clearly any potential residual gap in this spin-2 system should be very small and difficult to verify in finite-sized systems.

The challenge in explaining this anomalously critical-like behavior, together with the rarity of extended critical phases with symmetry emergence  motivated us to still explore higher-spin generalizations of this problem. We proposed a family of integer spin-$S$ Hamiltonians whose magnetic behaviors not only serve as examples for the limitations of the non-interacting spin-wave theory in 1D, but are all similarly related: they are all dominated by ferro-octupolar correlations and display dipolar and quadrupolar correlations with $2\pi/3$-periodicity. Higher-order multipolar correlations are naturally also present in these, either peaking at vanishing momentum or displaying period-three behavior. These results also signal how arbitrary spiral order is not favored in quantum magnetism, an effect that we assign to simple properties of angular-momentum summation for the generation of singlets. Amusingly, for spin $S=3$, the Hamiltonian we studied was found to be located in an interesting region of the parameter space with exact G$_2$ symmetry; this might be a promising starting point to search for G$_2$ criticality, of relevance for quantum computing applications.

Despite our results providing evidence against a conformal-invariant ground state in the spin-2 problem, an abnormal difficulty for DMRG convergence is present in this problem, which suggests either a very small gap or, indeed, the absence of one. If a gap exists, characterizing this phase would be an important step: not many candidates seem to remain to explain this beyond a trimerized state, but a 1D trimerized phase in an isotropic system seems as exotic as the critical phase proposed by Chen et al. As far as we can tell, if the system is indeed gapless, the SU$(3)_1$ state does seem to be the best candidate. Still, an analytical explanation for the emergence of a critical phase resisted our best efforts. To say the minimum, our numerical results indicate that an uncommon frustration mechanism is necessary, where degrees of freedom controlling octupolar and hexadecapolar fluctuations decouple from the dipolar and quadrupolar ones, with the former being spatially uniform while the latter ones giving rise to a quasi-SU$(3)_1$ fluid. Therefore, important questions remain. If the SU$(3)_1$ phase is there, how could it emerge? If it exists for spin-2, does it generalize to arbitrary integer spin? If this phase is not there, why does this system look so critical, and what is the nature and origin of these gapped ferro-octupolar phases instead? Finally, recent advances in tensor network methods have introduced methods that are efficient for simulation of gapless phases~\cite{CFT_numerics}; would it be possible to apply such methods to this present problem and improve the numerical data available, perhaps even including even including non-Abelian rotation invariance to these new methods? We leave these as questions for future considerations. 

\section{Acknowledgments}
We acknowledge insightful conversations with Fr\'ed\'eric Mila. For the SU(2) calculations we use an in-house custom code utilizing the High-Performance Tensor Transpose library \cite{hptt2017}, the Tensor Contraction Library \cite{tccg2016a}, the WIGXJPF library \cite{johansson2016}, and the FASTWIGXJ library \cite{Rasch:2003dr}. C. L. is supported by the QuEST scholorship at the University of British Columbia. For the early stages of this work, V.L.Q. acknowledges financial support from the High Magnetic Field Laboratory through the NSF Cooperative Agreement No. 1157490 and the State of Florida.

\appendix
\section{Other numerical results}\label{app:num}
\begin{figure*}
    \centering
    \includegraphics[width=\textwidth]{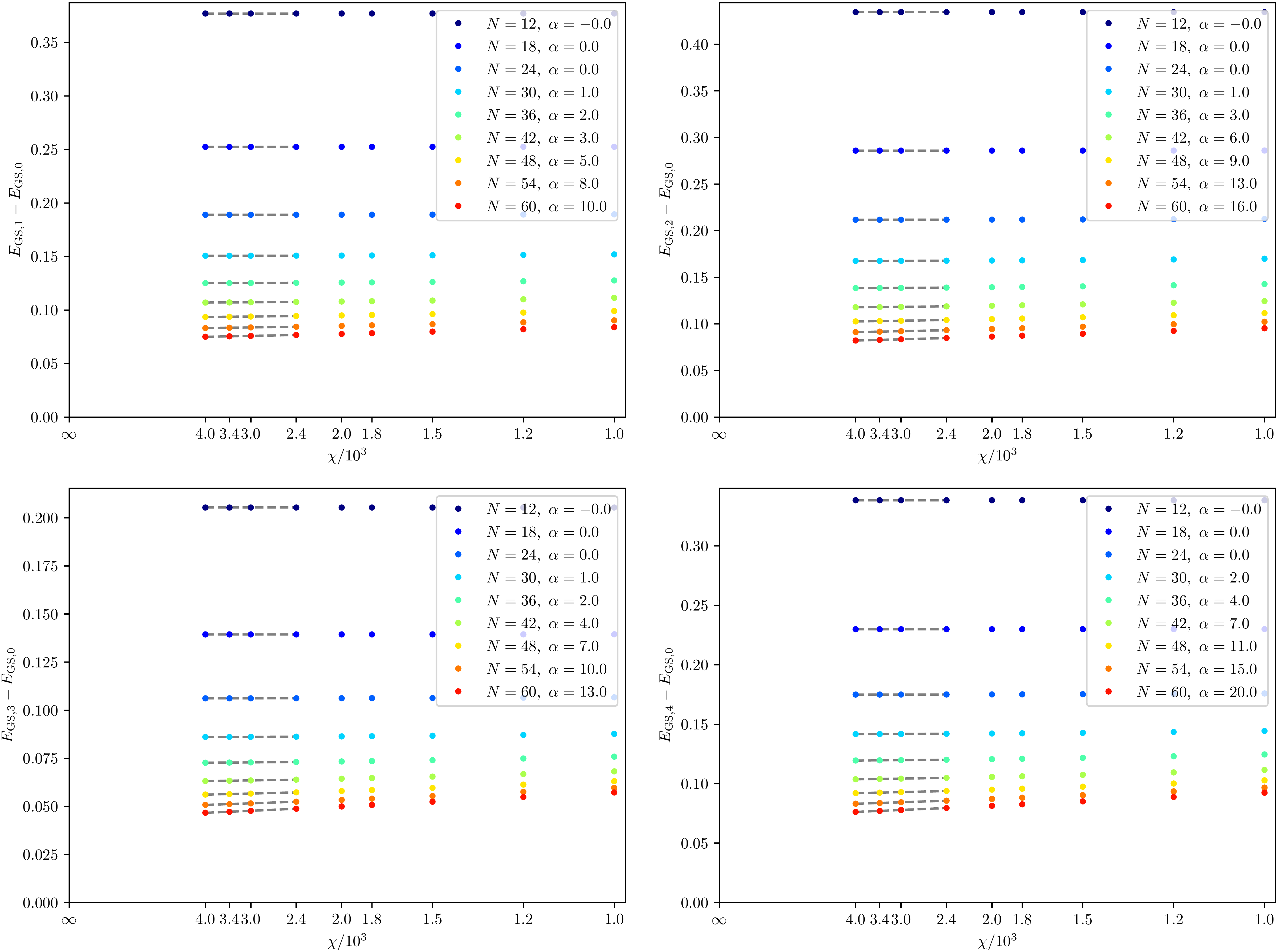}
    \caption{The energy gap $\Delta$ as a function of bond dimension $\chi$ between 1000 and 4000. A linear fit of $\Delta$ against $1/\chi$ on the largest four $\chi$'s gives a slope $\alpha$. }
    \label{fig:gap_chi}
\end{figure*}

Here we present the finite bond dimension scaling of the energy gap to indicate the convergence of the DMRG simulations. The energy gap $\Delta$ is obtained by subtracting the ground state energy in spin sectors of $S=1,..,4$ from that of the singlet. The results with bond dimension $\chi$ between 1000 and 4000 are shown in Fig.~\ref{fig:gap_chi}. For a given fixed system size, a linear fit of $\Delta$ against $1/\chi$ on the last four data sets, i.e.~$\chi=2400,3000,3400,4000$, gives an estimate of the convergence. Indeed, denoting the fitted slope as $\alpha$, we have a lower bound of $\Delta$,
\begin{equation}
\Delta_\infty\geq\Delta_{\chi_M}-\frac{\alpha}{\chi_M},
\end{equation}
where $\chi_M=4000$ is the largest available bond dimension and $\Delta_\infty$ is the infinite bond-dimension limit. We see that for $N\gtrsim48$, $\alpha\gtrsim10$, resulting in $\alpha/\chi_M\gtrsim0.0025$, or $\sim5\%$ of the value. With these system sizes, and convergence errors, we cannot rule out the possibility of this system being gapped.

\section{G$_2$ symmetry of the spin-3 model}\label{app:g2}
In this appendix we discuss the G$_2$ symmetry of the spin-3 Hamiltonian, $H_{3}=-\sum_{i}\mathcal{P}_{3}^{i}$. G$_2$ is the smallest exceptional semi-simple Lie group and also the automorphism group of the octonion algebra~\cite{Macfarlane2001}. We will focus on its corresponding Lie algebra $\mathfrak{g}_2$.

The Lie algebra has a two-dimensional Cartan subalgebra, for which we can choose the basis
\begin{equation}
\begin{split}
H_1=\frac{1}{\sqrt{14}}\mathrm{diag}\{-3,-2,-1,0,1,2,3\},\\
H_2=\frac{1}{\sqrt{42}}\mathrm{diag}\{-1,4,-5,0,5,-4,1\},
\end{split}
\end{equation}
which are proportional to $T_0^1$ and $T_0^5$, respectively. Then judiciously combining the eigenvectors of $[H_i,\boldsymbol{\cdot}]$ (a linear operator on $\mathfrak{sl}(7,\mathbb{C})$), we arrive at the generators. Here we adopt the standard notation with the two simple roots $\alpha_1=(\frac{3}{\sqrt{14}},\frac{1}{\sqrt{42}})$ and $\alpha_2=(-\frac{5}{\sqrt{14}},\frac{3}{\sqrt{42}})$. The root generators are 
\begin{equation}
\begin{split}
&E_1=\frac{1}{\sqrt{3}}
\begin{pmatrix}
0\\
0\\
0\\
\sqrt{2}\\
0 & -1\\
0 & 0 & -1\\
0 & 0 & 0 & \sqrt{2} & 0 & 0 & 0
\end{pmatrix},\\
&E_2=
\begin{pmatrix}
0 & 0 & 0 & 0 & 0 & 1 & 0\\
 &  &  &  &  &  & 1\\
 &  &  &  &  &  & 0\\
 &  &  &  &  &  & 0\\
 &  &  &  &  &  & 0\\
 &  &  &  &  &  & 0\\
 &  &  &  &  &  & 0\\
\end{pmatrix},\\
&E_{12}=[E_1,E_2], E_{112}=\frac{\sqrt{3}}{2}[E_1,E_{12}], \\
&E_{1112}=[E_1,E_{112}], E_{11122}=[E_2,E_{1112}],
\end{split}
\end{equation}
and their transposes. The set of two Cartan generators and the twelve root operators can be chosen as the generators of the group. With the matrix form of these generators, one can readily check that they commute with the Hamiltonian $H_3$ (for that, checking for a pair of sites is enough). In fact, the Hamiltonian can be expressed in term of the quadratic Casimir $C_2$ as
\begin{equation}
H_3=\sum_i \frac{1}{4}C_2^i-\frac{5}{16}(C_2^i)^2-\frac{3}{32}(C_2^i)^3,
\end{equation}
where $C_2$, invariant under group rotations, is
\begin{equation}
C_2^i=H_1^iH_1^{i+1}+H_2^iH_2^{i+1}+\sum_E E^i(E^{i+1})^\dagger.
\end{equation}
The sum over $E$ is the sum over all the twelve roots.

\bibliography{spin2SU3}

\end{document}